\documentclass[twocolumn,showpacs,floats,floatfix,superscriptaddress,aps,pra]{revtex4-1}
\usepackage{amsfonts}
\usepackage{amssymb}
\usepackage{amsmath}
\usepackage{graphicx}
\usepackage{bm}
\usepackage{color}
\usepackage{epsfig}
\usepackage{calc}
\usepackage{ifthen}
\usepackage{subfigure}
\usepackage{graphicx}
\usepackage{epstopdf}
\usepackage{epsfig}

\begin{document}

\title{Adiabatic following of terahertz surface plasmon-polaritons coupler via two waveguides structure}
\date{\today }

\begin{abstract}
Most recently, two remarkable papers [New J. Phys. 21, 113004 (2019); IEEE J. Sel. Top. Quantum Electron 27, 1 (2020)] propose broadband complete transfer terahertz (THz) surface plasmon-polaritons (SPPs) waveguide coupler by applying coherent quantum control - Stimulated Raman adiabatic passage (STIRAP). However, previous researches request three SPPs waveguides coupler. In this paper, we propose a new design of a broadband complete transfer THz SPPs coupler with an innovative structure of two waveguides by employing two state adiabatic following. In order to realize this design, we introduce the detuning parameter into the coupling equation of SPPs waveguides for the first time. We believe that this finding will improve the THz communication domain. 
\end{abstract}

\author{Wei Huang}
\affiliation{Guangxi Key Laboratory of Optoelectronic Information Processing, Guilin University of Electronic Technology, Guilin 541004, China}

\author{Weifang Yang}
\affiliation{Guangxi Key Laboratory of Optoelectronic Information Processing, Guilin University of Electronic Technology, Guilin 541004, China}

\author{Shan Yin}
\email{syin@guet.edu.cn}
\affiliation{Guangxi Key Laboratory of Optoelectronic Information Processing, Guilin University of Electronic Technology, Guilin 541004, China}

\author{Wentao Zhang}
\email{zhangwentao@guet.edu.cn}
\affiliation{Guangxi Key Laboratory of Optoelectronic Information Processing, Guilin University of Electronic Technology, Guilin 541004, China}

\maketitle



\section{Introduction}
Terahertz (THz) surface plasmon-polaritons (SPPs) waveguide coupler is the critical element device for THz communication \cite{Ye2018, Hasan2016, Liang2015} and detection \cite{Zhu2018}. The conventional THz SPPs coupler device \cite{Kumar2011, Ye2018, Ye2017, Zhang2018, Zhang2017, Liu2014, Maier2006} is made of two parallel coupling of SPPs waveguides and the disadvantages of those designs are that they only can operate at one single working frequency and the performance is very sensitive to geometric parameters. To solve those problems, most recently, we propose the broadband and robust complete transfer of THz SPPs waveguide coupler by employing coherent quantum control, named Stimulated-Raman-adiabatic-passage (STIRAP) \cite{Huang2019, Huang2020}. The coherent quantum control is a very robust method for the quantum system \cite{Vitanov2017, Huang2017} and currently, it is widely used in many classical systems, such as broadband optical waveguide coupler \cite{Ciret2012, Hristova2016, Huang20192}, nonlinear optics \cite{Erlich2019} and wireless energy transfer \cite{Rangelov2012, Huang20202}. 

The idea of the THz SPPs coupler device based on STIRAP is established by three special curve SPPs waveguides, which forces the coupling strengths of each two adjacent SPPs waveguides with satisfying STIRAP \cite{Huang2019, Huang2020}. However, those designs require three SPPs waveguides coupler and it brings complex fabrication process. In this paper, we propose a broadband complete transfer device with two SPPs waveguides coupler, which is realized by two-state adiabatic following. 

\begin{figure}[htbp]
	\centering
		\includegraphics[width=0.5\textwidth]{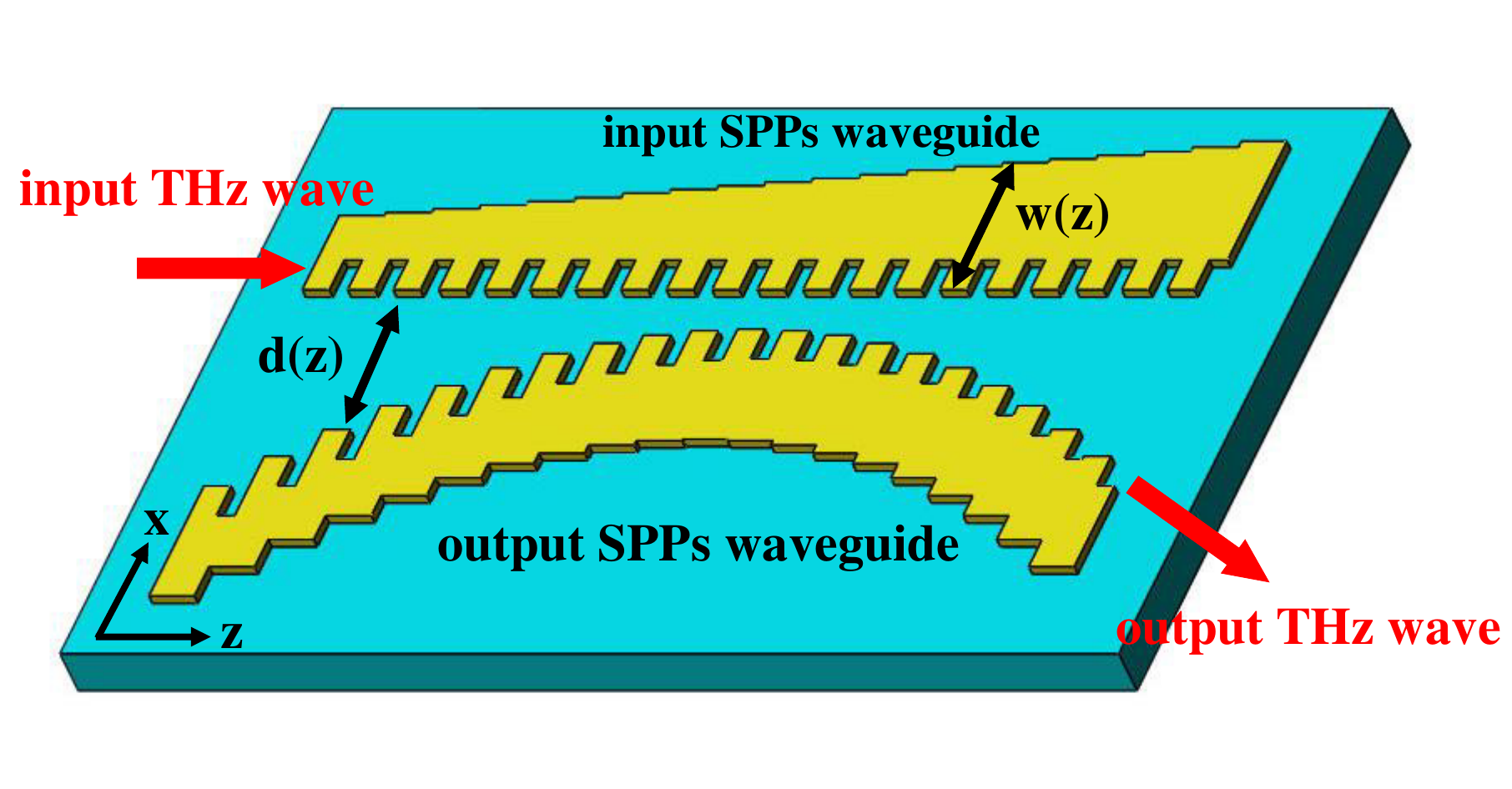}
	\caption{The schematic figure of our broadband device with two SPPs waveguides coupling. }
	\label{Fig1}
\end{figure}

The two-state adiabatic following is a well-known coherent quantum control technique, which requires tuning the coupling strength and detuning of the coupled equation. We have already well studied the coupling strength in the coupled equation of SPPs waveguides coupler \cite{Huang2019, Huang2020}. However, the detuning of the coupled equation is still an open problem and most recently, a remarkable paper demonstrates that the different width of SPPs waveguide can confine different prorogation constant of SPPs waveguide \cite{Gao2013}. Based on this idea, we introduce the detuning parameter into the coupled equation of THz SPPs waveguides for the first time in this paper. Therefore, we build up a complete Schrodinger equation for SPPs waveguides coupled equation. Subsequently, we can tune the geometrical parameters of the SPPs waveguides coupler to satisfy the coupling strength and detuning of the two-state adiabatic following. In this configuration, we can obtain the broadband complete transfer of THz SPPs waveguides coupler, as shown in Fig. \ref{Fig1}.

In this paper, we first introduce the detuning parameter of SPPs waveguides coupled equation and obtain the relationship between the geometrical parameter of SPPs waveguide and detuning parameter. Based on this finding, we can easily design the two SPPs waveguide coupler to achieve the broadband complete transfer device. In the section III, we run the numerical calculation of the coupled equation to demonstrate the complete transfer of SPPs energy. Besides, we use the full-wave simulation of our device with different frequencies of input THz waves to illustrate the broadband feature of our device.

\section{Adiabatic following of SPPs waveguides coupler}
It has already proofed that the coupled equation of SPPs waveguides is equivalent to Schrodinger equation \cite{Huang2019, Huang2018}, as written as,
\begin{equation}
i\dfrac{d}{d z}
\begin{bmatrix}
c_{1}(z) \\
c_{2}(z) 
\end{bmatrix}
= \begin{bmatrix}
\beta_1 (z) & \Omega (z) \\
\Omega (z) & \beta_2 (z)
\end{bmatrix} \begin{bmatrix}
c_{1} (z) \\
c_{2} (z)
\end{bmatrix},
\end{equation}
where $c_1$ and $c_2$ are the amplitudes of SPPs energy with respective to input and output SPPs waveguide. $\beta_1$ and $\beta_2$ are the propagation constants of input and output SPPs waveguides. $\Omega$ is the coupling strength between input and output SPPs waveguides. It is straight forward to obtain the detuning parameter $\Delta$, such as,
\begin{equation}
i\dfrac{d}{d z}
\begin{bmatrix}
c_{1}(z) \\
c_{2}(z) 
\end{bmatrix}
= \begin{bmatrix}
\Delta (z) & \Omega (z) \\
\Omega (z) & -\Delta (z)
\end{bmatrix} \begin{bmatrix}
c_{1} (z) \\
c_{2} (z)
\end{bmatrix},
\end{equation}
where the detuning $\Delta(z) = \frac{\beta_1(z) - \beta_2(z)}{2}$. 

We now write Eq. 2 in the so-called \textit{adiabatic basis} \cite{Shore2008,Vitanov2001} which is the eigenstates of the Hamiltonian,
\begin{equation}
i\frac{d}{dz}\left[
\begin{array}{c}
a_{1} \\
a_{2}%
\end{array}%
\right] =\left[
\begin{array}{cc}
-\sqrt{\Omega ^{2}+\Delta ^{2}} & -i\dot{\vartheta} \\
i\dot{\vartheta} & \sqrt{\Omega ^{2}+\Delta^{2}}%
\end{array}%
\right] \left[
\begin{array}{c}
a_{1} \\
a_{2}%
\end{array}%
\right] ,  \label{Adiabatic basis}
\end{equation}%
where $\tan 2\vartheta =\frac{\Omega}{\Delta}$. 
The connection between the amplitudes $c_{1}$ and $c_{N}$ and the adiabatic states $a_{1}$ and $a_{2}$ is given by
\begin{subequations}
\label{adiabatic states}
\begin{align}
a_{1}& =c_{1}\cos \vartheta -c_{2}\sin \vartheta , \\
a_{2}& =c_{1}\sin \vartheta +c_{2}\cos \vartheta .
\end{align}%
\end{subequations}

When the condition meets adiabatic condition \cite{Shore2008,Vitanov2001}, shown as,
\begin{equation}
\left\vert \dot{\vartheta}\right\vert \ll \left( \Omega ^{2}+\Delta^{2}\right) ^{1/2}.
\end{equation}%
Mathematically, adiabatic evolution means that the non-diagonal terms in Eq. (\ref{Adiabatic
basis}) are small compared to the diagonal terms and can be neglected. $|a_{1}|^{2}$ and $|a_{2}|^{2} $ remain constant \cite{Shore2008,Vitanov2001}. 
Then takes the explicit form
\begin{equation}
\frac{2\left( \Omega^{2}+\Delta^{2}\right) ^{3/2}}{\left\vert
\Delta \dot{\Omega} -\Omega \dot{\Delta}\right\vert }%
\gg 1, \label{adiabatic condition}
\end{equation}
where $\dot{\Omega}$ and $\dot{\Delta}$ are derivative with $z$ of $\Omega$ and $\Delta$. 
For our case the detuning $\Delta$ change from some positive to some negative value, in this course the mixing angle $\vartheta $
changes from $\pi /2$ to $0$. With the SPPs energy initially in the input waveguide, the system stays adiabatically in state $a_{2}$, so that the energy transfer to the output waveguide when $\vartheta $ approaches zero. We employ a typical configuration for coupling strength and detuning, which makes the Gaussian shape for coupling strength and linear shape for detuning. 

\begin{figure}[htbp]
	\centering
		\includegraphics[width=0.5\textwidth]{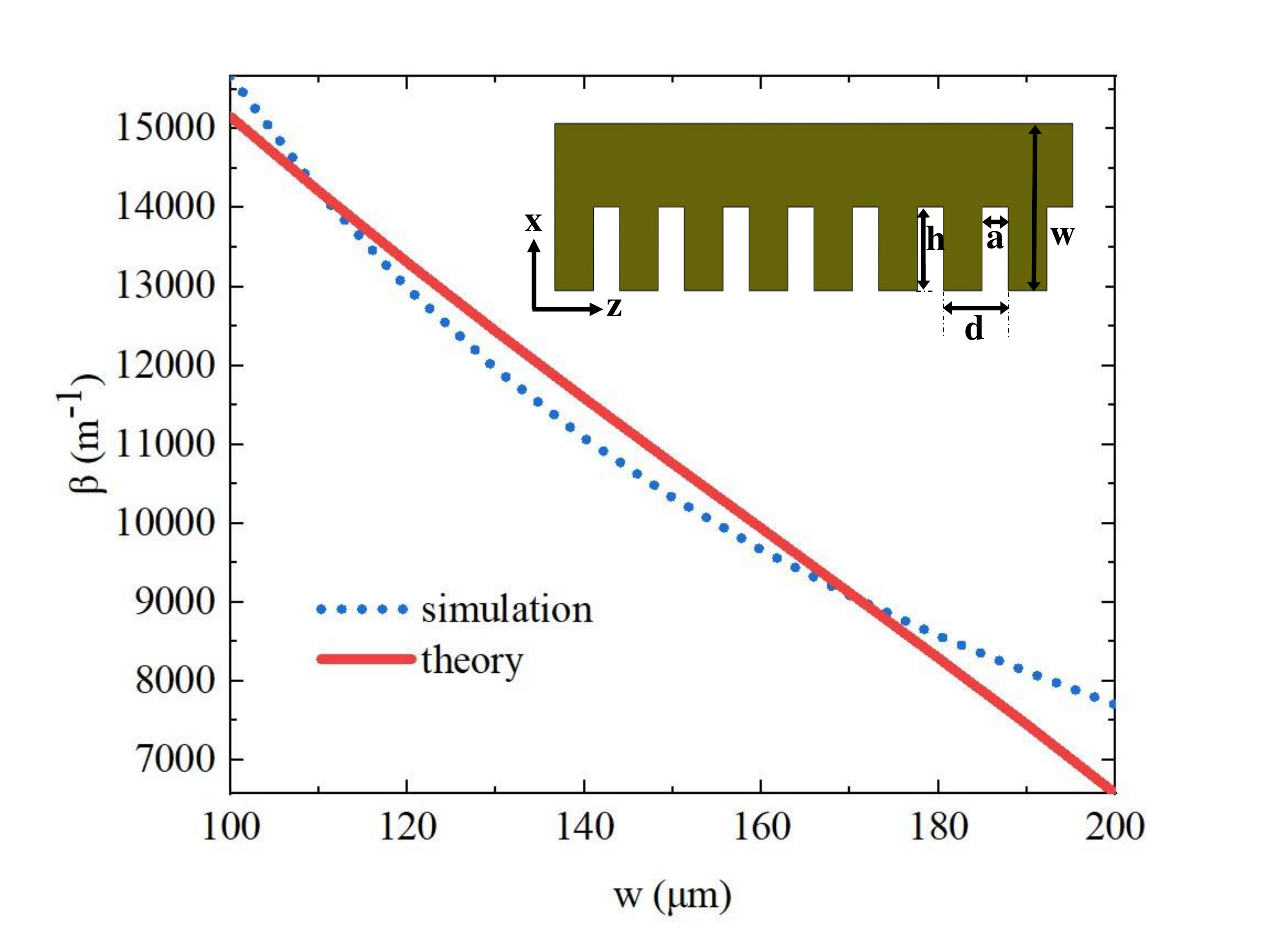}
	\caption{The schematic figure of our broadband device with two SPPs waveguides coupling. }
	\label{Fig2}
\end{figure}

\begin{figure}[htbp]
	\centering
		\includegraphics[width=0.45\textwidth]{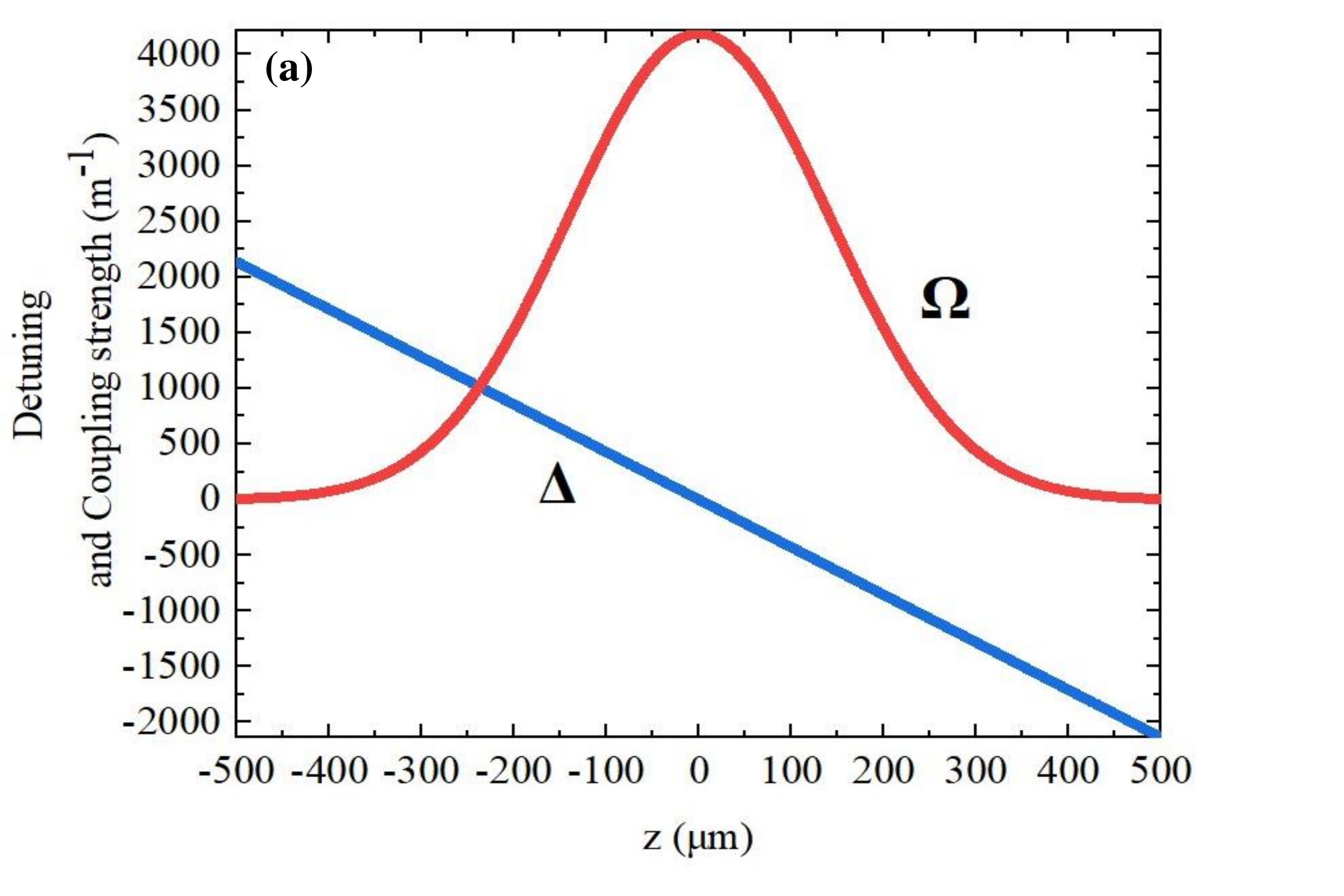}
		\includegraphics[width=0.45\textwidth]{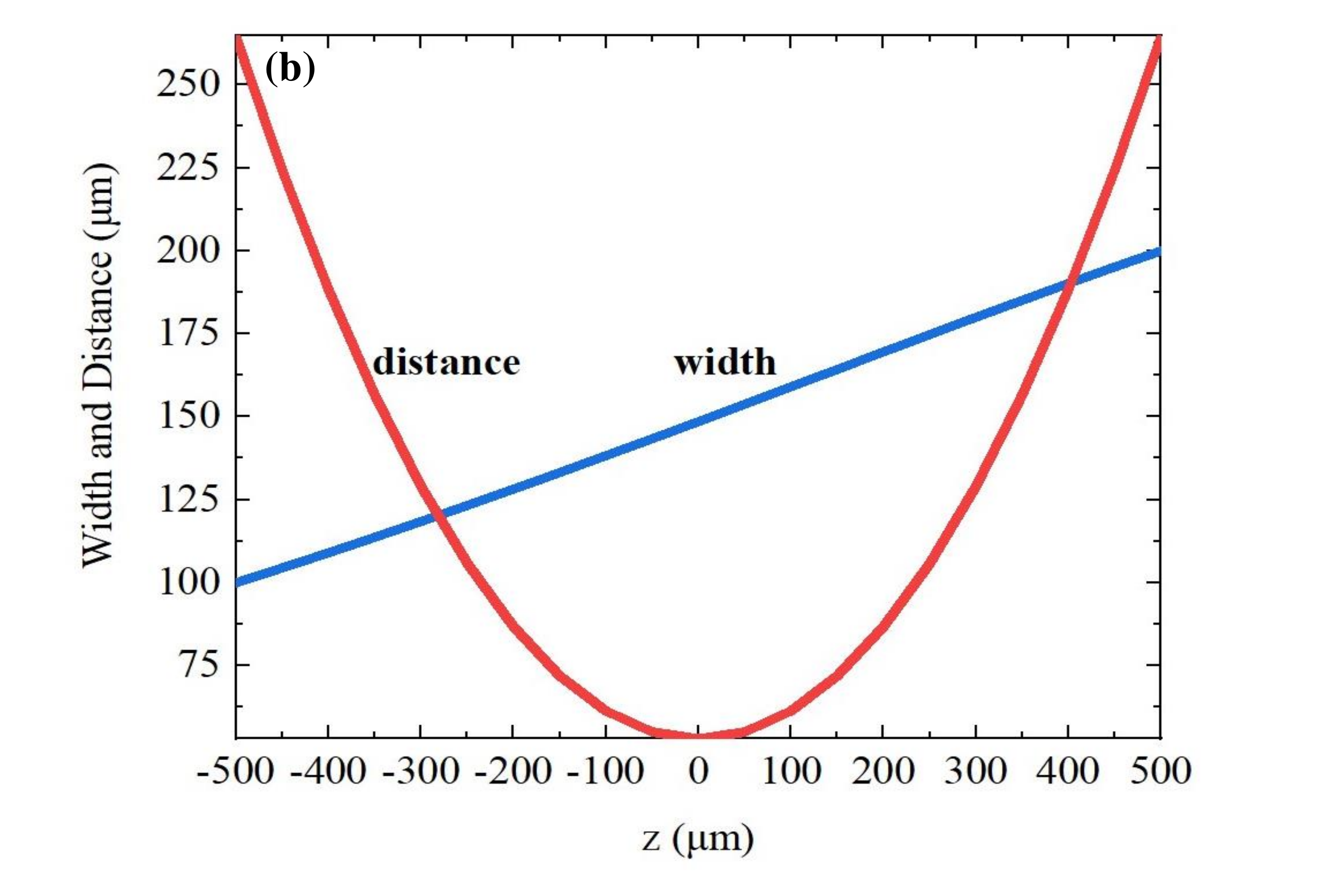}
		\includegraphics[width=0.45\textwidth]{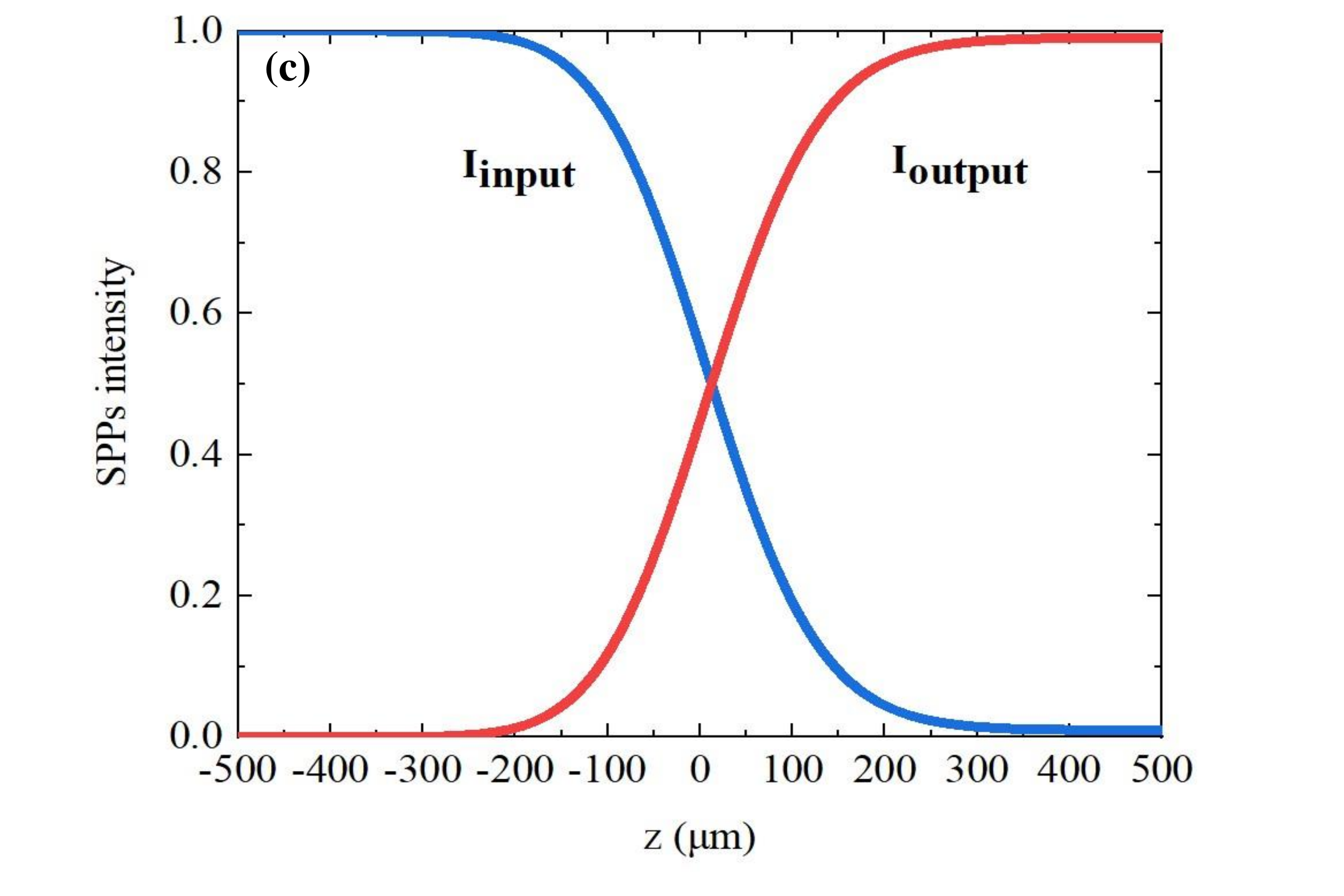}
	\caption{(a) The functions of coupling strength $\Omega (z)$ and detuning $\Delta (z)$ along with $z$, (b) corresponding to geometrical parameters of SPPs waveguides and (c) corresponding to the evolution of SPPs energy transfer by numerical calculations for Eq. 2.}
	\label{Fig3}
\end{figure}

\begin{figure*}[htbp]
		\subfigure{\includegraphics[width=0.45\textwidth]{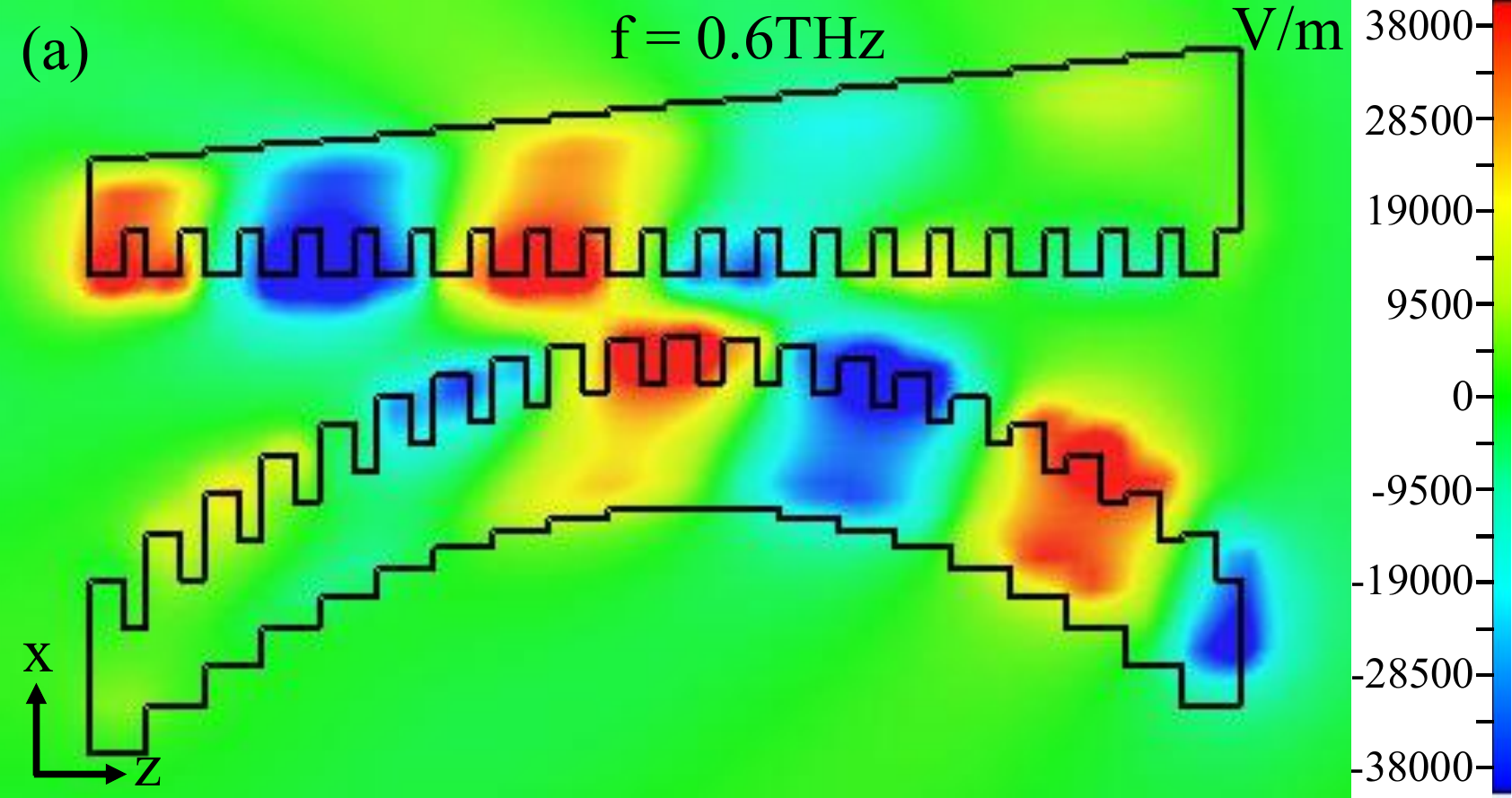}}
		\subfigure{\includegraphics[width=0.45\textwidth]{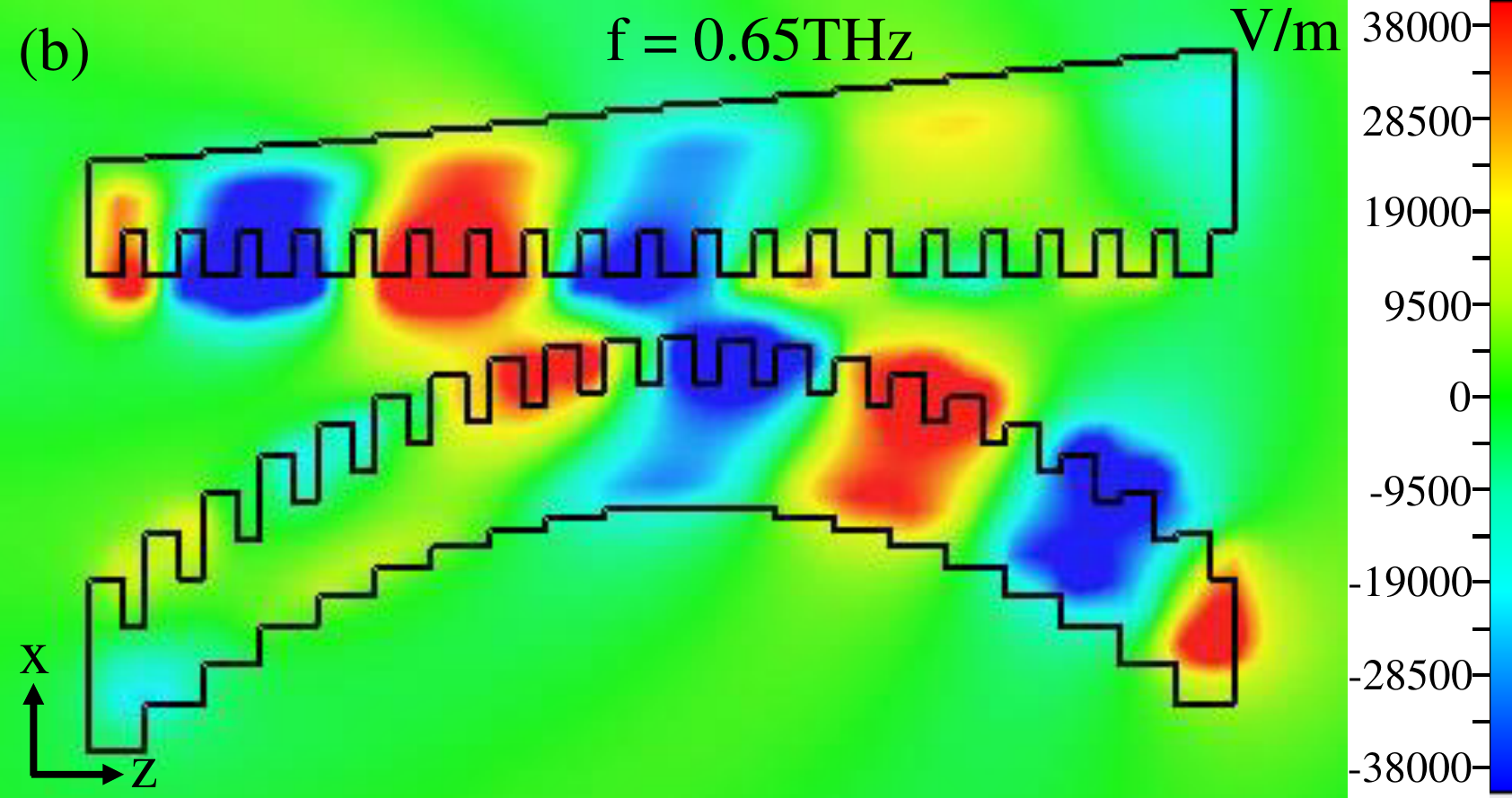}}
		\subfigure{\includegraphics[width=0.45\textwidth]{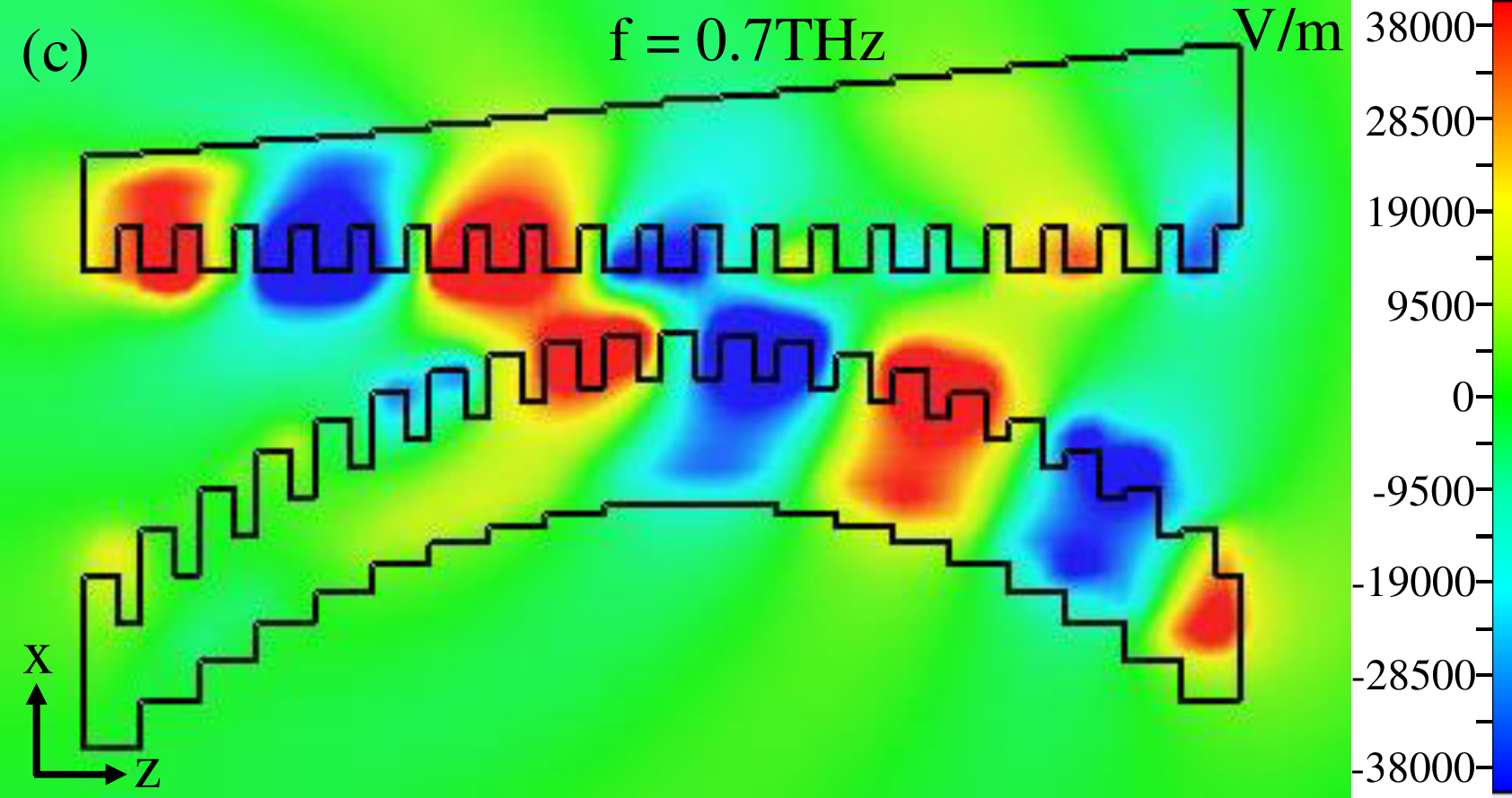}}
		\subfigure{\includegraphics[width=0.45\textwidth]{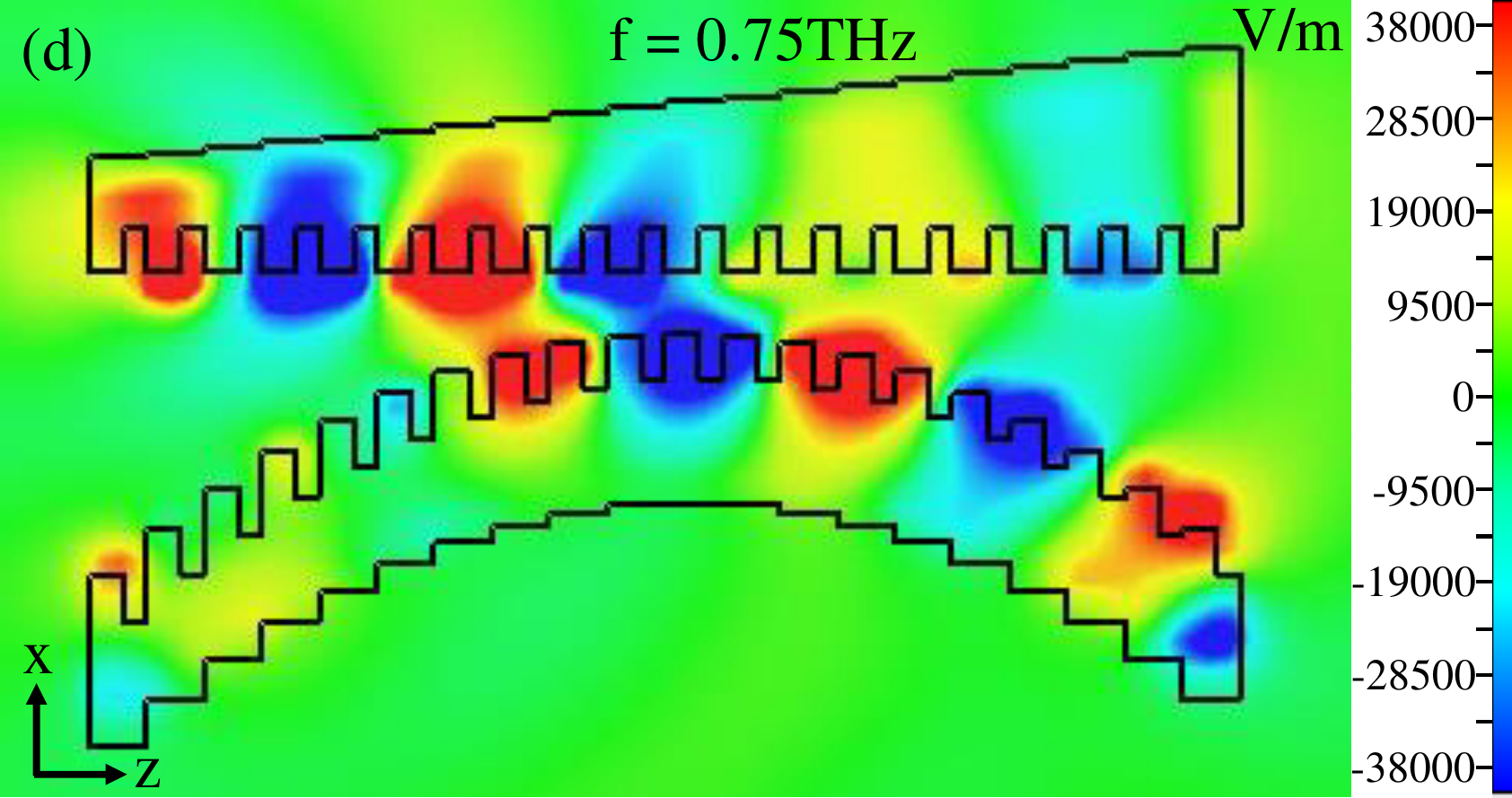}}
		\subfigure{\includegraphics[width=0.45\textwidth]{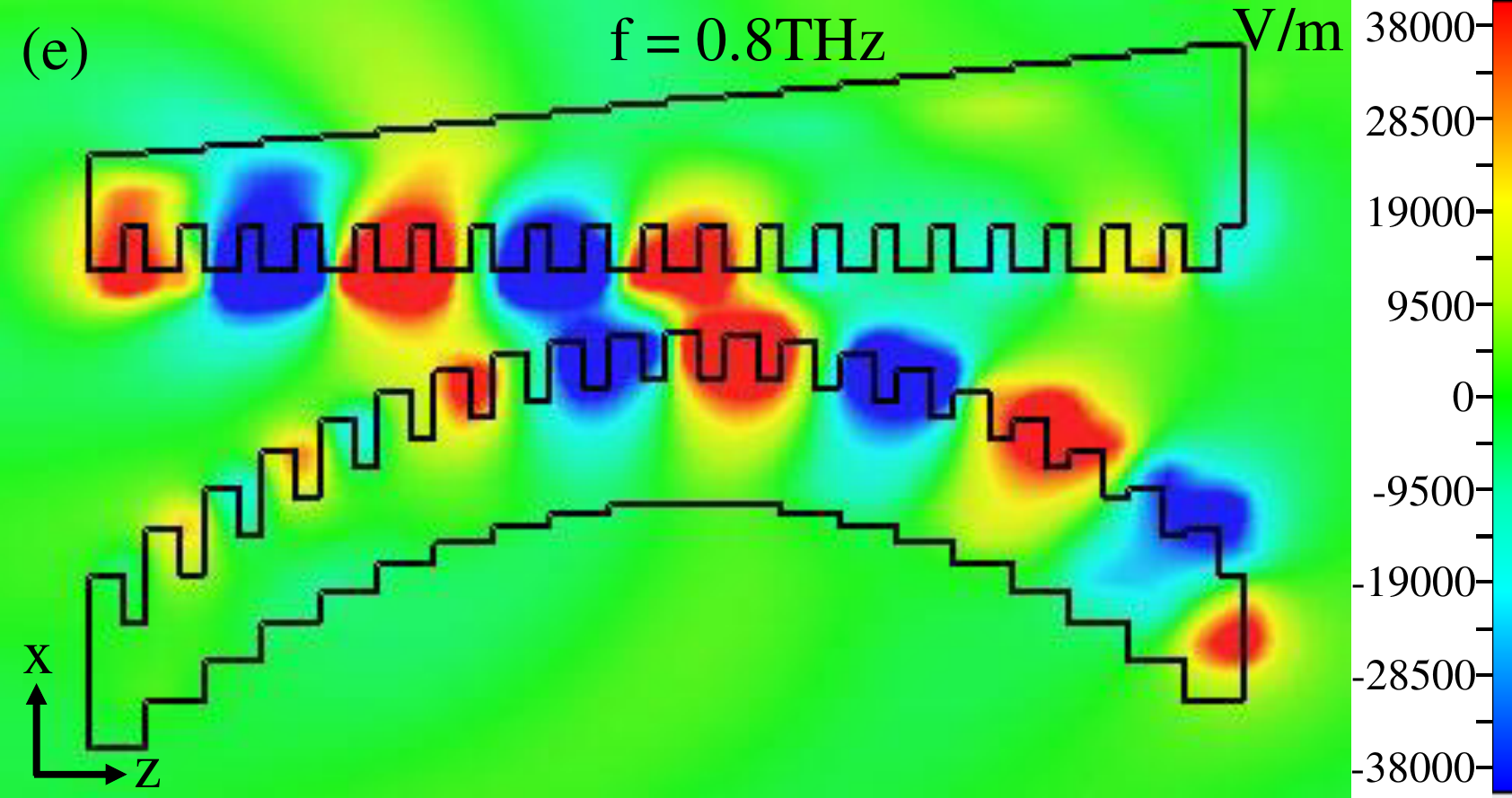}}
		\subfigure{\includegraphics[width=0.45\textwidth]{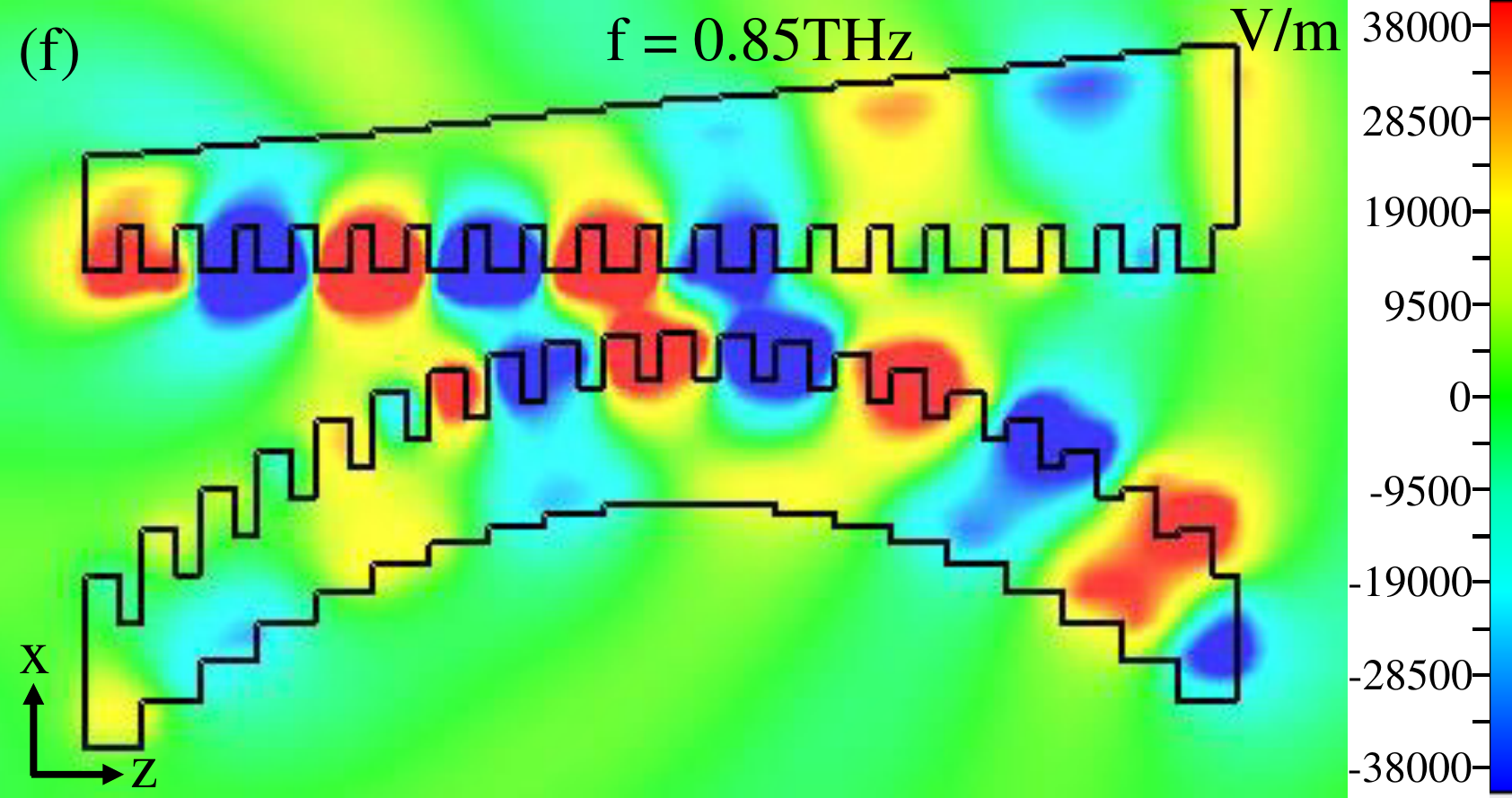}}
		\subfigure{\includegraphics[width=0.45\textwidth]{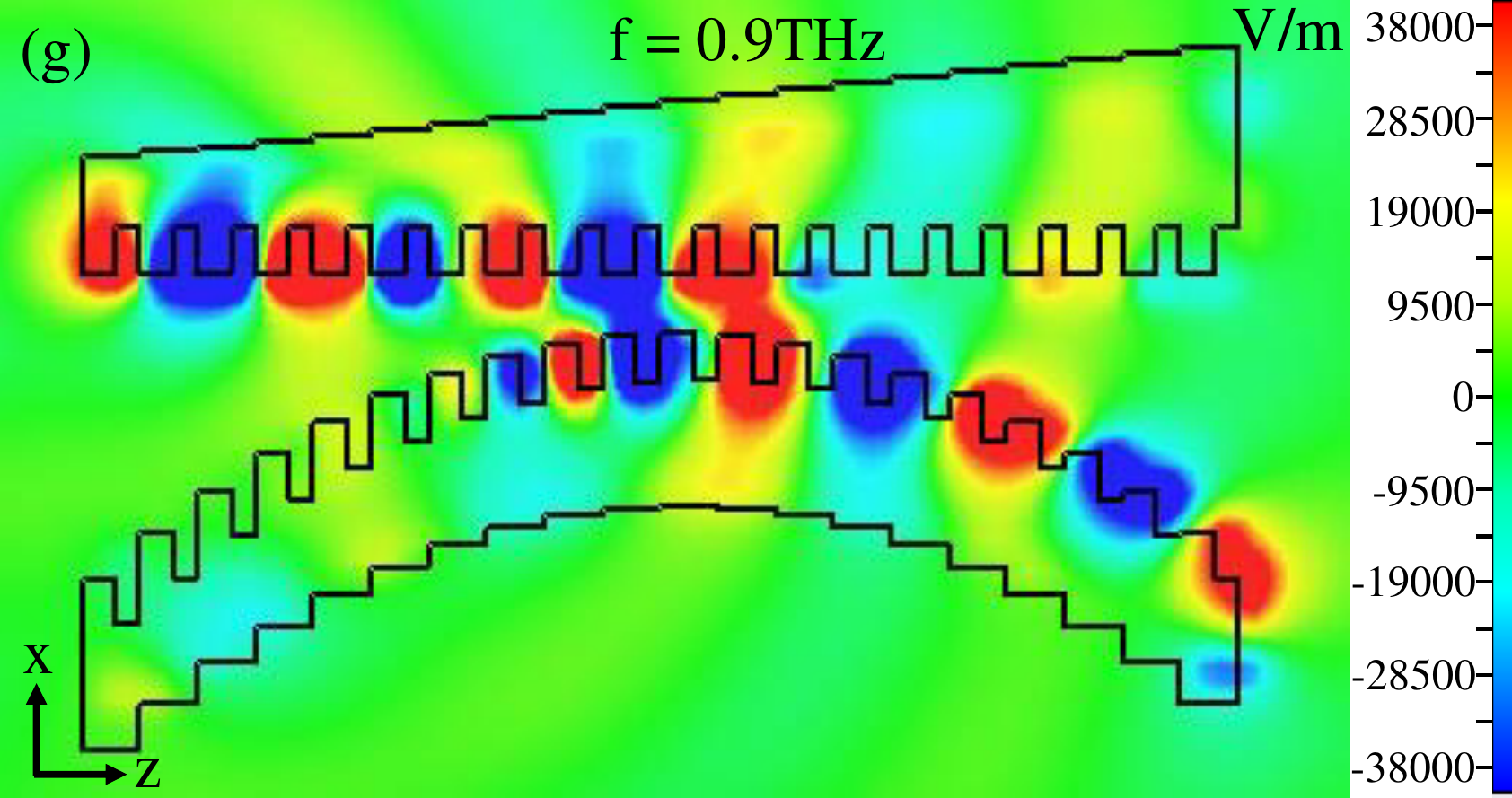}}
	\caption{The full-wave simulations of our device with input THz wave frequencies. }
	\label{Fig4}
\end{figure*}

\section{Device design}
In order to tune the coupling strength and detuning, we obtain the relationship between geometrical parameters of SPPs waveguide and coupling strength (and detuning) in this section.
In this paper, we use a periodical structure for THz SPPs waveguide \cite{Shen2013, ShenXP2013} with fixed h = 40 $\mu m$, a = 20 $\mu m$ and d = 50 $\mu m$, as shown in the small figure of Fig. \ref{Fig2}. 
For the coupling strength, it is well-known about tuning the coupling strength $\Omega(z)$ with changing distance between two SPPs waveguides, which has an exponential relationship with the distance between two SPPs waveguides \cite{Huang2019, Huang2020} based on the coupled mode theory (CMT). 

For the detuning, we obtain that the width $w$ of SPPs waveguide can remarkably tune the propagation constant $\beta$ of the SPPs waveguide, as shown in Fig. \ref{Fig2}. The blue line of Fig. \ref{Fig2} is the results based on the simulation of dispersion equation with varying the width $w$ of SPPs waveguide. Our theory is based on the propagation constant on the plane metal, which is $\beta = \frac{\omega}{c}\sqrt{\frac{\epsilon_d \epsilon_m}{\epsilon_d+\epsilon_m}}$ \cite{Takayama2017}, where $\omega$ is the frequency of input THz wave and $c$ is the light speed. $\epsilon_d$ and $\epsilon_m$ are the permittivities of substrate and metal. Then we can modify the propagation constant $\beta$ with our structure,
\begin{equation}
\beta = \frac{\omega}{c}\sqrt{\frac{\epsilon_d \epsilon'_m}{\epsilon_d+\epsilon'_m}}, 
\end{equation}
where $\epsilon'_m$ is the effective permittivity which can be controlled by the width $w$ for $\epsilon'_m = A \frac{[(d-a)w+a(w-h)]\epsilon_m + [ah] \epsilon_d}{wd}$, where $A$ is the fitting number as shown in red line of Fig. \ref{Fig2}. 

From the result of Fig. \ref{Fig2}, thus we vary the width $w$ of SPPs waveguide to control the detuning $\Delta(z)$. 
Based on the adiabatic following theory and tuning coupling strength $\Omega(z)$ and detuning $\Delta(z)$, we can give an example of set of detuning $\Omega(z)$ and $\Delta(z)$, given as
\begin{equation}
\begin{aligned}
\Omega(z) &=  \Omega_0 \text{exp}(-(\frac{z^2}{b^2}));  \\ 
\Delta(z) &= \frac{2 \Delta_0 z}{-L},
\end{aligned}
\end{equation}
where $\Omega_0$ is the maximum coupling strength with minimum distance between two SPPs waveguides and $\Delta_0$ is maximum detuning based on the difference between the minimum and maximum width of SPPs waveguide. We take the parameters b = 200 $\mu m$ and L is the length of our device from -500 $\mu m$ to 500 $\mu m$. The functions of coupling strength and detuning are shown in Fig. \ref{Fig3} (a). 

Therefore, we vary the distance between two SPPs waveguide to change the coupling strength $\Omega(z)$ (see the red line of Fig. \ref{Fig3} (b)) and vary the width of input SPPs waveguide to change the detuning $\Delta (z)$ (see the blue line of Fig. \ref{Fig3} (b)). In this configuration of the distance of two SPPs waveguides and the width of the input SPPs waveguide, the coupling strength and detuning satisfy the adiabatic following (Fig. \ref{Fig3} (a)). Therefore, we run the numerical calculation of Eq. 2 with the functions of coupling strength $\Omega (z)$ and detuning $\Delta (z)$, as shown in Fig. \ref{Fig3} (c). From the result, it is very easy to find that the SPPs energy can complete flow from input to output SPPs waveguide.

\section{Full-wave simulations}
Due to the complex boundary condition of SPPs waveguides, it is very tough to analytically and numerically calculate the evolution of SPPs energy based on Eq. 2. Therefore, in order to demonstrate the broadband feature of our device, we employ the full-wave simulations of our device with different input THz waves but keeping the same structure of our device from 0.6 THz to 0.9 THz, as shown in Fig. \ref{Fig4}. As we can see from the results (Fig. \ref{Fig4}), our device has very good performance with different input THz waves. Indeed, in order to change the detuning $\Delta (z)$ of our device, we should vary the width of the input SPPs waveguide. It indeed brings some perturbations of SPPs energy with spreading the whole width of SPPs waveguide, which leads some SPPs energy on upper input SPPs waveguide not to couple and transfer to output SPPs waveguide. This is the reason why some energy of SPPs still remains within the input SPPs waveguide. However, based on our full-wave simulations, our device can perform relatively good performance with complete transfer SPPs energy from input to output SPPs waveguide.

\section{Conclusion}
In this paper, we obtain the detuning parameter of SPPs waveguides coupled equation which is built up a full model for two-level quantum system for the first time. By tuning the detuning and coupling strength via the geometrical parameters of our device, we propose a broadband and complete transfer two SPPs waveguide coupler by employing two-state adiabatic following. Our device not only can maintain the broadband feature, but also reduce the fabrication complexity and cost. We believer that this finding will improve the research domain in THz communication and THz imaging.

\section*{Acknowledgements}
This work acknowledges funding from National Key Research and Development Program of China (2019YFB2203901); National Science and Technology Major Project (grant no: 2017ZX02101007-003); National Natural Science Foundation of China (grant no: 61565004; 61965005; 61975038; 62005059). The Science and Technology Program of Guangxi Province (grant no: 2018AD19058). W.H. acknowledges funding from Guangxi oversea 100 talent project; W.Z. acknowledges funding from Guangxi distinguished expert project.


\begin{thebibliography}{99}
\bibitem{Ye2018} Ye, Longfang, et al. "Super subwavelength guiding and rejecting of terahertz spoof SPPs enabled by planar plasmonic waveguides and notch filters based on spiral-shaped units." Journal of Lightwave Technology 36.20 (2018): 4988-4994.

\bibitem{Hasan2016} Hasan, Mehdi, et al. "Graphene terahertz devices for communications applications." Nano Communication Networks 10 (2016): 68-78.

\bibitem{Liang2015} Liang, Yuan, et al. "On-chip sub-terahertz surface plasmon polariton transmission lines in CMOS." Scientific reports 5.1 (2015): 1-13.

\bibitem{Zhu2018} Zhu, Jiaqi, et al. "Terahertz imaging sensor based on the strong coupling of surface plasmon polaritons between PVDF and graphene." Sensors and Actuators B: Chemical 264 (2018): 398-403.

\bibitem{Kumar2011} Kumar, Gagan, et al. "Planar plasmonic terahertz waveguides based on periodically corrugated metal films." New Journal of Physics 13.3 (2011): 033024.

\bibitem{Ye2018} Ye, Longfang, et al. "Super Subwavelength Guiding and Rejecting of Terahertz Spoof SPPs Enabled by Planar Plasmonic Waveguides and Notch Filters Based on Spiral-Shaped Units." Journal of Lightwave Technology 36.20 (2018): 4988-4994.

\bibitem{Ye2017} Ye, Longfang, et al. "Plasmonic waveguide with folded stubs for highly confined terahertz propagation and concentration." Optics Express 25.2 (2017): 898-906.

\bibitem{Zhang2018} Zhang, Ying, et al. "Terahertz spoof surface-plasmon-polariton subwavelength waveguide." Photonics Research 6.1 (2018): 18-23.

\bibitem{Zhang2017} Zhang, Ying, et al. "Terahertz surface plasmon polariton waveguiding with periodic metallic cylinders." Optics Express 25.13 (2017): 14397-14405.

\bibitem{Liu2014} Liu, Xiaoyong, et al. "Planar surface plasmonic waveguide devices based on symmetric corrugated thin film structures." Optics express 22.17 (2014): 20107-20116.

\bibitem{Maier2006} Maier, Stefan A. , et al. "Terahertz Surface Plasmon-Polariton Propagation and Focusing on Periodically Corrugated Metal Wires." Physical Review Letters 97.17(2006):176805.

\bibitem{Huang2019} Huang, Wei, et al. "Robust and broadband integrated terahertz coupler conducted with adiabatic following." New Journal of Physics 21.11 (2019): 113004.

\bibitem{Huang2020} Huang, Wei, et al. "Quantum Engineering Enables Broadband and Robust Terahertz Surface Plasmon-Polaritons Coupler." IEEE Journal of Selected Topics in Quantum Electronics 27.2 (2020): 1-7.

\bibitem{Gao2013} Gao, Xi, et al. "Ultrathin dual-band surface plasmonic polariton waveguide and frequency splitter in microwave frequencies." Applied Physics Letters 102.15 (2013): 151912.

\bibitem{Shen2013} Shen, xiaopeng, et al. "Conformal surface plasmons propagating on ultrathin and flexible films." Proc Natl Acad Sci U S A 110.1 (2013): 40-45.

\bibitem{ShenXP2013} Shen, xiaopeng, et al. "Planar plasmonic metamaterial on a thin film with nearly zero thickness." Applied Physics Letters 102.21 (2013): 211909.

\bibitem{Takayama2017} Takayama O., Bogdanov AA., et al. "Photonic surface waves on metamaterial interfaces." Journal of Physics-Condensed Matter 29.46 (2017):463001.
 
\bibitem{Vitanov2017} Vitanov, Nikolay V., et al. "Stimulated Raman adiabatic passage in physics, chemistry, and beyond." Reviews of Modern Physics 89.1 (2017): 015006.

\bibitem{Huang2017} Huang, Wei, et al. "Adiabatic following for a three-state quantum system." Optics Communications 382 (2017): 196-200.

\bibitem{Ciret2012} Ciret, Charles, et al. "Planar achromatic multiple beam splitter by adiabatic light transfer." Optics letters 37.18 (2012): 3789-3791.

\bibitem{Hristova2016} Hristova, Hristina S., et al. "Adiabatic three-waveguide coupler." Physical Review A 93.3 (2016): 033802.

\bibitem{Huang20192} Huang, Wei, Lay-Kee Ang, and Elica Kyoseva. "Shortcut to adiabatic light transfer in waveguide couplers with a sign flip in the phase mismatch." Journal of Physics D: Applied Physics 53.3 (2019): 035104.

\bibitem{Erlich2019} Erlich, Yonathan, et al. "Robust, efficient, and broadband SHG of ultrashort pulses in composite crystals." Optics letters 44.15 (2019): 3837-3840.

\bibitem{Rangelov2012} Rangelov, Andon A., and Nikolay V. Vitanov. "Mid-range adiabatic wireless energy transfer via a mediator coil." Annals of Physics 327.9 (2012): 2245-2250.

\bibitem{Huang20202} Huang, Wei, et al. "Long-distance adiabatic wireless energy transfer via multiple coils coupling." Results in Physics 19 (2020): 103478.

\bibitem{Huang2018} Huang, Wei, et al. "Adiabatic control of surface plasmon-polaritons in a 3-layers graphene curved configuration." Carbon 127 (2018): 187-192.

\bibitem{Shore2008} B. W. Shore, Acta Phys. Slovaca \textbf{58}, 243 (2008).

\bibitem{Vitanov2001} N. V. Vitanov, M. Fleischhauer, B. W. Shore, and K.

\end{thebibliography}
\end{document}